# Enhanced Photovoltaic Performance of Perovskite Solar Cells by Co-Doped Spinel Nickel Cobaltite Hole Transporting Layer

*Apostolos Ioakeimidis[1], Ioannis T. Papadas[1], Dimitris Tsikritzis[1], Gerasimos S. Armatas[2], Stella Kennou[3], Stelios A. Choulis[1]\**

[1] Molecular Electronics and Photonics Research Unit, Department of Mechanical Engineering and Materials Science and Engineering, Cyprus University of Technology, Limassol, 3603, Cyprus
[2] Department of Materials Science and Technology, University of Crete, Heraklion 71003, Greece
[3] Department of Chemical Engineering, University of Patras, 26504, Patras, Greece

*Corresponding Author: Prof. Stelios A. Choulis
E-mail:stelios.choulis@cut.ac.cy

## Abstract

Solution combustion synthesized hole transport layer (HTL) of spinel nickel cobaltite ($NiCo_2O_4$) incorporating 3% Cu – 2% Li were fabricated using doctor-blading technique for planar inverted perovskite solar cells (PVSCs). PVSCs incorporating 3% Cu - 2% Li-doped $NiCo_2O_4$ shown an increase in Jsc and Voc device performance parameters compared to unmodified $NiCo_2O_4$, leading to PCE of 16.5%. X-ray photoelectron spectroscopy measurements revealed the tendency of Cu cations to replace preferably the surface Ni atoms changing the surface stoichiometry of $NiCo_2O_4$ inducing a cathodic polarization. Ultraviolet photoelectron spectroscopy measurements unveiled the increase of the ionization potential by 0.1 eV for co-doped $NiCo_2O_4$ film compare to unmodified $NiCo_2O_4$-based HTL. We attribute the enhanced PCE of inverted PVSCs presented due to improved hole extraction properties of 3% Cu - 2% Li $NiCo_2O_4$ HTL.

Perovskite solar cells have amazed with an incredibly fast power conversion efficiency (PCE) improvement, going from 3.8% in 2009[1] to over 20% in 2018.[2–5] A lot of parameters have been investigated to increase the performance and reliability of the devices such as the element composition[6–11] and preparation method of perovskite[12–21], device configuration[22–26], materials and preparation conditions of hole/electron transporting layers[27–33].

Regarding the investigation of functional hole transporting layers (HTLs), a wide variety of organic and inorganic materials have been implemented to improve hole extraction, with some of the latter's advantage to be the wide optical band gap (thus high transparency in the visible range) and superior hole mobility, while they can be solution processed. Some promising inorganic HTLs are $NiO_x$,[34] $Cu:NiO_x$[35–37], $CuO_x$[38–40], $CuI$[41], $CuSCN$,[42] $CuGaO_2$,[43] $CuCrO_2$[44]. Recently, we have reported combustion synthesis of monodispersed spinel $NiCo_2O_4$ nanoparticles of ~4 nm diameter forming a compact layer with electrical conductivity of ~4 S/cm. The developed films were applied as an efficient and reliable HTL for inverted structure perovskite solar cells (PVSCs) using 230 nm thick perovskite layer.[45] In order to increase the PCE of the devices a thicker perovskite layer is needed. The enhancement in light absorption leads to an increase in photogenerated carriers which accumulate at the perovskite/HTL interface (accumulation zone) and subsequently collected by the contact[46]. Thus, HTL with enhanced hole collection capability are required to increase the PCE. A common method to enhance HTL charge collection efficiency is by incorporation of intentional defects through extrinsic doping. This process can induce a higher electrical conductivity as well as better energy level alignment of HTL with the perovskite active layer.[47–51] For example, recently a co-doping strategy of $NiO_x$ with Cu/Li or Li/Mg elements has been successfully applied to enhance the PCE of PVSC.[52,53]

In this paper, we report the use of solution combustion synthesized $NiCo_2O_4$ co-doped with 3 mol% Cu and 2 mol% Li (3% Cu – 2% Li) as efficient HTL to increase the performance of inverted PVSCs. Initially, $NiCo_2O_4$ film doped with 5 mol% Cu was incorporated as HTL in PVSC exhibiting an increased Voc. However, the Jsc of the corresponding PVSC has declined significantly compare to unmodified $NiCo_2O_4$-based PVSC due to lower electrical conductivity. We show that an increase in the electrical conductivity can be achieved, by 3% Cu and 2% Li co-doping of the $NiCo_2O_4$-HTL resulting in PVSCs with enhancement on both Voc and Jsc compare to $NiCo_2O_4$-HTL based PVSCs.

X-ray photoelectron spectroscopy (XPS) investigation on co-doped $NiCo_2O_4$-HTL showed a decrease on the Ni/Co atomic ratio compared to unmodified $NiCo_2O_4$-HTL, indicating the preferable surface substitution of nickel by copper cations which has been previously reported that it induces a cathodic polarization.[54] As a result, an increase of the ionization potential by 0.1 eV was observed for 3% Cu – 2% Li $NiCo_2O_4$-HTLs compare to stoichiometric $NiCo_2O_4$-HTL using ultraviolet photoelectron spectroscopy (UPS). The increased performance of the reported PVSCs could be attributed to the cathodic polarization potential and thus better hole collection efficient of the 3% Cu – 2% Li $NiCo_2O_4$ layer.

The PVSCs under investigation were prepared on top of glass/Indium Tin Oxide (ITO)/$NiCo_2O_4$ for the different doping types processed as described in detail at the supplementary section. The perovskite solution was prepared 30 min prior spin coating by mixing $Pb(CH_3CO_2)_2 \cdot 3H_2O$:methylamonium iodide (1:3) at 40 wt% in dimethylformamide (DMF) with the addition of 1.5% mole of MABr (methylammonium bromide). Briefly, ~350 nm perovskite active layer was spin-coated on top of each substrate followed by 50 nm spin-coated $PC_{70}BM$ (serving as the electron selective contact) and 100 nm thermally deposited Al. More details for the materials and processing conditions can be found within the supplementary section.

Figure 1(b) demonstrates the current density - voltage (J – V) measurements under 1 sun simulated illumination for the PVSCs using $NiCo_2O_4$ with different doping types and the extracted photovoltaic parameters are shown in Table I. Pristine $NiCo_2O_4$ HTL based PVSCs shows a considerable lower Voc (0.88 V) but higher Jsc (18.25 mA/cm$^2$) compare to 5% Cu-doped $NiCo_2O_4$ HTL based PVSCs (Voc = 1.03 V, Jsc = 14.89 mA/cm$^2$), while FF is almost similar (72.3 % and 73 %)), delivering a PCE of 11.61% and 11.02%, respectively. Importantly, PVSCs incorporating 3% Cu – 2% Li $NiCo_2O_4$ HTL exhibits both higher Voc (1.05 V) and Jsc (21.05 mA/cm$^2$) as well as a slightly higher FF (74.8 %) compare to previous devices, delivering a PCE of 16.54 %.

Table I. J -V extracted parameters of PVSC using 15 nm undoped, 5% Cu and 3% Cu – 2% Li $NiCo_2O_4$ as HTL.

| HTL | Voc (V) | Jsc (mA/cm$^2$) | FF (%) | PCE (%) |
|---|---|---|---|---|
| $NiCo_2O_4$ | 0.88 | 18.25 | 72.3 | 11.61 |
| 5% Cu | 1.03 | 14.89 | 73 | 11.02 |
| 3 % Cu - 2 % Li | 1.05 | 21.05 | 74.8 | 16.54 |

In order to investigate the reduced photocurrent of 5% Cu-doped $NiCo_2O_4$, we first excluded any possible optical losses induced by the doping. Figure 1(a) demonstrates the transmittance of ~15 nm-thick $NiCo_2O_4$ layer on glass/ITO. It is obvious that the difference on transmittance is negligible for all films under study, where the extracted

Tauc-plot (Fig. S1) for direct transitions [ ( α.E )$^2$ = A.( E - Eg) ] show similar optical band gaps (Egs). Further, the similar morphology in all types of $NiCo_2O_4$ films was confirmed excluding, also, differences in electrical losses related to films quality (e.g. shunting current). Fig. S2 and Fig. S3 illustrates the AFM topography images of (a) 5% Cu and (b) 3% Cu -2% Li $NiCo_2O_4$ films fabricated on quartz and glass/ITO substrates, while Fig. S3(c) illustrates the topography of the ITO underlayer. In both cases the films exhibit similar roughness between them ( 0.7 - 0.8 nm for quartz and 2.9 - 3.0 nm for glass/ITO substrate) comparable to the ones measured for the pristine $NiCo_2O_4$ films, affirming the similar quality of different types of $NiCo_2O_4$ films.[45]

Thus, electrical characterization of PVSC were performed using Electro impedance spectroscopy (EIS) measurements under illumination and zero bias on the previous described PVSCs configurations. As it is observed at Figure 1(c) all the spectra show the characteristic two frequency response, where the first arc (higher frequencies) is ascribed to charge transfer resistance ($R_{ct}$) while the second larger arc (lower frequencies) at the charge carrier recombination resistance ($R_{rec}$).[55,56] PVSC incorporating 3% Cu – 2% Li $NiCo_2O_4$-HTL exhibits higher $R_{rec}$ compare to unmodified $NiCo_2O_4$-HTL based PVSCs, while shows lower $R_{ct}$ [Fig 1(c), inset)]compare to both unmodified and 5% Cu-doped $NiCo_2O_4$-HTL based PVSCs due to higher electrical conductivity of the 3% Cu – 2% Li $NiCo_2O_4$ layer, as it was also confirmed by four point probe conductivity measurements summarized within Table II. Jung-Hee Kim *et. al.* have also reported increase in the electrical conductivity of spinel nickel cobaltite by introduction of Li.[57]

Table II. Room-temperature four-point probe extracted values of undoped $NiCo_2O_4$ and 5% Cu and 3% Cu – 2% Li $NiCo_2O_4$ films.

| NiCo$_2$O$_4$ | Conductivity (S/cm) |
|---|---|
| undoped | 4.00 |
| 5% Cu | 1.87 |
| 3% Cu - 2% Li | 4.85 |

Additional Mott-Schottky (Fig. 1(d)) measurements were carried out on devices sweeping from higher to lower voltage under dark conditions. The crossing of the curves at $1/C^2 = 0$ is attributed to the flat band potential of the device.[58,59] 5% Cu and 3% Cu – 2% Li NiCo$_2$O$_4$ -HTL based PVSCs show a higher built-in potential compare to unmodified NiCo$_2$O$_4$-HTL based PVSCs which is consistent with the increased Voc value achieved for the 5% Cu and 3% Cu – 2% Li NiCo2O4 -HTL based PVSCs..

Further investigation of the charge carrier recombination dynamics was conducted to elucidate the enhanced device performance of 3% Cu – 2% Li doped compare to undoped NiCo$_2$O$_4$-HTL based PVSCs. We first exclude any difference on the perovskite film morphology. AFM topography images (Fig. S4) of perovskite surface revealed similar surface roughness (12.5 ± 0.4 nm) and grain sizes (ca. 110-123 nm) as shown within the paper supplementary information (Fig. S5), indicating that PVSCs under study comprise similar morphology within the active layer. Moreover, Voc - light intensity measurements were performed to investigate the recombination mechanism within PVSCs under study. According to simplified Shockley Reed Hall recombination model, the slope between logarithmic light intensity and Voc must be equal to 2kT/q for trap-assisted and kT/q for trap-free recombination[60–64]. As shown in Fig.2(a), the Voc - light intensity curves scale equal to kT/q, implying that a trap-free recombination mechanism is dominant for all the PVSCs within this paper. Thus, steady state photoluminescence (PL) measurements (Fig.2(b)) are adequate to evaluate the degree of charge recombination at each configuration. The PL intensity for undoped

NiCo$_2$O$_4$-HTL is much higher compared to 3% Cu – 2% Li NiCo$_2$O$_4$-HTL, implying that a much higher number of electron-hole pairs recombine for the case of the undoped HTL justifying the lower PCE of the corresponding undoped NiCo2O4-HTL based PVSCs. The experimental results presented indicate that 3% Cu – 2% Li NiCo$_2$O$_4$ HTL transfers and collects hole charges more efficient than the undoped NiCo$_2$O$_4$ HTL.

A deeper material properties and device physics investigation was performed to better understand the origin of the enhanced hole collection properties for 3% Cu – 2% Li NiCo$_2$O$_4$. Structural characterization with X-ray diffraction (XRD) on the corresponding NiCo$_2$O$_4$ samples (Fig.S6) matched the cubic face-centered lattice structure of NiCo$_2$O$_4$ (PDF#20-0781), implying single-crystalline structure. X-ray photoelectron spectroscopy (XPS) measurements were also performed on doped and undoped NiCo$_2$O$_4$ HTLs. The Co 2p spectrum (Fig.S7) was best fitted by using two spin-orbit doublets for the tetrahedral Co$^{+2}$ and octahedral Co$^{+3}$ oxidation states and with two shake-up satellites located at the higher binding energy (BE) side of the main peaks. The peak located around 779.7 eV can be attributed to the octahedral Co$^{+3}$ observed in Co$_3$O$_4$,[65] while the higher binging energy peak around 780.9 eV can be assigned to the tetrahedral Co$^{+2}$ similar to CoO.[66] The spectrum of the Ni 2p$_{3/2}$ region was fitted using three components (Fig. S8). The peak at 854.3 eV corresponds to Ni$^{+2}$ ions, while that at 856.0 eV is attributed to Ni$^{+3}$.[65,67] The shake-up satellite at around 861.8 eV was fitted considering one broad line. For Cu doped films the Cu 2p spectra were recoded and are displayed in Fig. S9. The Cu 2p doublet is well resolved. The Cu 2p3/2 peak at 934.6 eV and the satellite at higher binding energies indicate that Cu is oxidized and can be identified as Cu$^{+2}$ ions in octahedral coordination.[54,67,68] The intensity of Cu 2p3/2 peak for the 3% Cu -2% Li NiCo$_2$O$_4$ is low and the satellite structure is not resolved. Nevertheless, the peak is located at BEs around 934.6 eV, thus even for lower concentration of Cu there are $Cu^{+2}$ ions. The table III summarize the Ni:Co atomic ratio values obtained from the processing of the reported XPS spectra. The surface sensitivity of XPS and material precursor stoichiometry reveals that a small excess of Ni ions is identified at the surface of the undoped $NiCo_2O_4$ as the XPS calculated Ni:Co ratio is 0.55. For 5% Cu doped and 3% Cu -2% Li $NiCo_2O_4$ a decrease at Ni:Co ratio confirms the deficiency of Ni ion at the surface, resulting in 0.43 and 0.45 ratios, respectively, which have been preferentially replaced by the Cu ions. These findings agree with previous reported results of A.C. Tavaresa *et al.*[54] where the introduction of Cu replace surface Ni ions at the $NiCo_2O_4$ electrodes, which indeed induces a similar effect to cathodic polarization (downshift of the energy bands).

Table III. Nickel to cobalt (Ni:Co) ratio extracted by XPS analysis of undoped, 5% Cu, and 3% Cu – 2% Li $NiCo_2O_4$ samples.

| Ratio | $NiCo_2O_4$ | 5 % Cu | 3 % Cu -2 % Li |
|---|---|---|---|
| Ni:Co | 0.55 | 0.43 | 0.45 |

Additional ultraviolet photoelectron spectroscopy (UPS) measurements were also performed on doped and undoped $NiCo_2O_4$ films to determine the energy levels. Figure 3(b) displays the UPS spectra of the valence band region near the Fermi level. The valence band maximum (VBM) for the $NiCo_2O_4$ was found at 0.2 eV bellow the Fermi level, while it is shifted to higher binding energies (~0.3 eV) when the $NiCo_2O_4$ is doped with 5% Cu and 3% Cu-2% Li. Figure 1(a) shows the high binding energy region of UPS spectra, where the high energy cut-off region is used to determine the work function ($\Phi$) of the interface. $\Phi$ for all films of $NiCo_2O_4$ was found at 5.1 eV and the ionization potential was calculated by adding the values of $\Phi$ and VBM. Thus, ionization potentials were found to be ~5.3 eV for the undoped $NiCo_2O_4$ and ~5.4 eV for 5% Cu and 3% Cu-2% Li NiCo$_2$O$_4$ HTLs. A schematic representation of PVSC energy band levels applying different types of NiCo$_2$O$_4$ layers is illustrated in Fig3(c) where the calculated values from the UPS and the E$_g$ (~2.3 eV) from UV-Vis optical absorption spectra were utilized. To summarize, the induced cathodic polarization of the 3% Cu-2% Li NiCo$_2$O$_4$ HTL increase the built-in potential of the corresponding PVSCs as shown using Mott Schottky measurements and reduce the charge recombination losses (as inferred from EIS results presented) due to better hole extraction (as inferred from PL measurements presented), giving rise to both Voc and Jsc compare to undoped NiCo$_2$O$_4$ HTL based PVSC.

In conclusion, we report the doping of co-doped NiCo$_2$O$_4$ with 5% Cu and 3% Cu – 2% Li to increase the PCE of inverted PVSC using a 350 nm Pb(CH3CO2)2.3H2O:methylamonium iodide (1:3) based perovskite formulation. 5% Cu doping increase the Voc of the corresponding PVSC but decrease the Jsc compare to undoped NiCo$_2$O$_4$ PVSC due to lower electrical conductivity. To overcome this effect 3% Cu – 2% Li co-doping was applied on solution combustion synthesized NiCo$_2$O$_4$-HTLinducing an increase on electrical conductivity resulting in inverted PVSCs with lower charge transfer resistance compare to 5% Cu doped NiCo$_2$O$_4$ -HTL based PVSCs and higher charge recombination resistance compare to undoped NiCo$_2$O$_4$-HTL based PVSCs. Mott Schottky measurements showed the higher built-in potential of the Cu doped NiCo$_2$O$_4$ PVSC while PL studies confirmed the better hole extraction of 3% Cu – 2% Li NiCo$_2$O$_4$-HTL/perovskite active layer interface. Further investigation for the origin of this enhancement was performed by XPS measurements on the co-doped and undoped NiCo$_2$O$_4$ revealing the tendency of Cu ions to replace preferably the surface Ni ions of NiCo$_2$O$_4$ changing the surface stoichiometry of Ni:Co which induces a cathodic polarization effect. UPS measurements reveled the increase of the ionization potential by 0.1 eV for the 3% Cu – 2% Li NiCo$_2$O$_4$ HTLs compared to undoped NiCo$_2$O$_4$-HTL a parameter which improves hole carrier extraction properties for the 3% Cu – 2% Li NiCo$_2$O$_4$-HTL based PVSCs reported. As a result, inverted PVSCs containing 3% Cu – 2% Li co-doped NiCo$_2$O$_4$ HTL shown an increased PCE of 16.54% compare to undoped NiCo$_2$O$_4$-HTL based PVSCs with PCE of 11.61%.

## Supplementary Material

Supplementary information includes details for the proposed hole transporting layer materials, processing of perovskite films and fabrication of perovskite devices. Additional information for materials optical characterization, surface topography, XRD and UPS/XPS measurements and analysis of the experimental results is included.

## Acknowledgments

This project received funding from the European Research Council (ERC) under the European Union's Horizon 2020 research and innovation program (grant agreement No. 647311). The authors would like to thank Associate Professor Gregorios Itskos for fruitful discussions.

## References

[1] A. Kojima, K. Teshima, Y. Shirai, and T. Miyasaka, J. Am. Chem. Soc. **131**, 6050 (2009).

[2] X. Zhu, D. Yang, R. Yang, B. Yang, Z. Yang, X. Ren, J. Zhang, J. Niu, J. Feng, and S. (Frank) Liu, Nanoscale **9**, 12316 (2017).

[3] F. Zhang, Z. Wang, H. Zhu, N. Pellet, J. Luo, C. Yi, X. Liu, H. Liu, S. Wang, X. Li, Y. Xiao, S.M. Zakeeruddin, D. Bi, and M. Grätzel, Nano Energy **41**, 469 (2017).

[4] W.S. Yang, B.-W. Park, E.H. Jung, N.J. Jeon, Y.C. Kim, D.U. Lee, S.S. Shin, J. Seo, E.K. Kim, J.H. Noh, and S. Il Seok, Science (80-. ). **356**, 1376 (2017).

[5] T. Singh and T. Miyasaka, Adv. Energy Mater. **8**, 1700677 (2018).

[6] N.J. Jeon, J.H. Noh, Y.C. Kim, W.S. Yang, S. Ryu, and S. Il Seok, Nat. Mater. **13**, 1 (2014).

[7] D.P. McMeekin, G. Sadoughi, W. Rehman, G.E. Eperon, M. Saliba, M.T. Horantner, A. Haghighirad, N. Sakai, L. Korte, B. Rech, M.B. Johnston, L.M. Herz, and H.J. Snaith, Science (80-. ). **351**, 151 (2016).

[8] F. Hao, C.C. Stoumpos, D.H. Cao, R.P.H. Chang, and M.G. Kanatzidis, Nat. Photonics **8**, 489 (2014).

[9] G.E. Eperon, V.M. Burlakov, P. Docampo, A. Goriely, and H.J. Snaith, Adv. Funct. Mater. **24**, 151 (2014).

[10] M. Saliba, T. Matsui, K. Domanski, J.Y. Seo, A. Ummadisingu, S.M. Zakeeruddin, J.P. Correa-Baena, W.R. Tress, A. Abate, A. Hagfeldt, and M. Grï¿½tzel, Science (80-. ). **354**, 206 (2016).

[11] N.J. Jeon, J.H. Noh, W.S. Yang, Y.C. Kim, S. Ryu, J. Seo, and S. Il Seok, Nature **517**, 476 (2015).

[12] M. Liu, M.B. Johnston, and H.J. Snaith, Nature **501**, 395 (2013).

[13] D. Liu and T.L. Kelly, Nat. Photonics **8**, 133 (2014).

[14] J. Burschka, N. Pellet, S.J. Moon, R. Humphry-Baker, P. Gao, M.K. Nazeeruddin, and M. Grätzel, Nature **499**, 316 (2013).

[15] Q. Chen, H. Zhou, Z. Hong, S. Luo, H.S. Duan, H.H. Wang, Y. Liu, G. Li, and Y.

Yang, J. Am. Chem. Soc. **136**, 622 (2014).

[16] D. Bi, S.-J. Moon, L. Häggman, G. Boschloo, L. Yang, E.M.J. Johansson, M.K. Nazeeruddin, M. Grätzel, and A. Hagfeldt, RSC Adv. **3**, 18762 (2013).

[17] F. Huang, Y. Dkhissi, W. Huang, M. Xiao, I. Benesperi, S. Rubanov, Y. Zhu, X. Lin, L. Jiang, Y. Zhou, A. Gray-Weale, J. Etheridge, C.R. McNeill, R.A. Caruso, U. Bach, L. Spiccia, and Y.B. Cheng, Nano Energy **10**, 10 (2014).

[18] X. Li, D. Bi, C. Yi, J.-D. Decoppet, J. Luo, S.M. Zakeeruddin, A. Hagfeldt, and M. Gratzel, Science (80-. ). **353**, 58 (2016).

[19] C.W. Chen, H.W. Kang, S.Y. Hsiao, P.F. Yang, K.M. Chiang, and H.W. Lin, Adv. Mater. **26**, 6647 (2014).

[20] A. Ioakeimidis, C. Christodoulou, M. Lux-Steiner, and K. Fostiropoulos, J. Solid State Chem. **244**, 20 (2016).

[21] H.A. Abbas, R. Kottokkaran, B. Ganapathy, M. Samiee, L. Zhang, A. Kitahara, M. Noack, and V.L. Dalal, APL Mater. **3**, (2015).

[22] P. Docampo, J.M. Ball, M. Darwich, G.E. Eperon, and H.J. Snaith, Nat. Commun. **4**, 2761 (2013).

[23] B. Conings, L. Baeten, C. De Dobbelaere, J. D'Haen, J. Manca, and H.G. Boyen, Adv. Mater. **26**, 2041 (2014).

[24] H.P. Zhou, Q. Chen, G. Li, S. Luo, T.B. Song, H.S. Duan, Z.R. Hong, J.B. You, Y.S. Liu, and Y. Yang, Science (80-. ). **345**, 542 (2014).

[25] J.-H. Im, C.-R. Lee, J.-W. Lee, S.-W. Park, and N.-G. Park, Nanoscale **3**, 4088 (2011).

[26] H.-S. Kim, C.-R. Lee, J.-H. Im, K.-B. Lee, T. Moehl, A. Marchioro, S.-J. Moon, R. Humphry-Baker, J.-H. Yum, J.E. Moser, M. Grätzel, and N.-G. Park, Sci. Rep. **2**, 591 (2012).

[27] L.E. Polander, P. Pahner, M. Schwarze, M. Saalfrank, C. Koerner, and K. Leo, APL Mater. **2**, 081503 (2014).

[28] O. Malinkiewicz, A. Yella, Y.H. Lee, G.M.M. Espallargas, M. Graetzel, M.K. Nazeeruddin, and H.J. Bolink, Nat. Photonics **8**, 128 (2014).

[29] M. Ye, C. He, J. Iocozzia, X. Liu, X. Cui, X. Meng, M. Rager, X. Hong, X. Liu, and Z. Lin, J. Phys. D. Appl. Phys. **50**, 373002 (2017).

[30] V. Zardetto, B.L. Williams, A. Perrotta, F. Di Giacomo, M.A. Verheijen, R. Andriessen, W.M.M. Kessels, and M. Creatore, Sustain. Energy Fuels **1**, 30 (2017).

[31] Z.H. Bakr, Q. Wali, A. Fakharuddin, L. Schmidt-Mende, T.M. Brown, and R. Jose, Nano Energy **34**, 271 (2017).

[32] T.Y. Wen, S. Yang, P.F. Liu, L.J. Tang, H.W. Qiao, X. Chen, X.H. Yang, Y. Hou, and H.G. Yang, Adv. Energy Mater. **8**, 1703143 (2018).

[33] D. Ouyang, J. Xiao, F. Ye, Z. Huang, H. Zhang, L. Zhu, J. Cheng, and W.C.H. Choy, Adv. Energy Mater. **8**, 1702722 (2018).

[34] L.J. Tang, X. Chen, T.Y. Wen, S. Yang, J.J. Zhao, H.W. Qiao, Y. Hou, and H.G. Yang, Chem. - A Eur. J. **24**, 2845 (2018).

[35] F. Galatopoulos, A. Savva, I.T. Papadas, and S.A. Choulis, APL Mater. **5**, (2017).

[36] K. Yao, F. Li, Q. He, X. Wang, Y. Jiang, H. Huang, and A.K.Y. Jen, Nano Energy **40**, 155 (2017).

[37] W. Chen, Y. Wu, J. Fan, A.B. Djurišić, F. Liu, H.W. Tam, A. Ng, C. Surya, W.K. Chan, D. Wang, and Z.B. He, Adv. Energy Mater. **1703519**, 1 (2018).

[38] W. Sun, Y. Li, S. Ye, H. Rao, W. Yan, H. Peng, Y. Li, Z. Liu, S. Wang, Z. Chen, L. Xiao, Z. Bian, and C. Huang, Nanoscale **8**, 10806 (2016).

[39] H. Rao, S. Ye, W. Sun, W. Yan, Y. Li, H. Peng, Z. Liu, Z. Bian, Y. Li, and C. Huang, Nano Energy **27**, 51 (2016).

[40] A. Savva, I.T. Papadas, D. Tsikritzis, G.S. Armatas, S. Kennou, and S.A. Choulis, J. Mater. Chem. A **5**, 20381 (2017).

[41] J.A. Christians, R.C.M. Fung, and P. V. Kamat, J. Am. Chem. Soc. **136**, 758 (2014).

[42] N. Wijeyasinghe, A. Regoutz, F. Eisner, T. Du, L. Tsetseris, Y.-H. Lin, H. Faber, P. Pattanasattayavong, J. Li, F. Yan, M.A. McLachlan, D.J. Payne, M. Heeney, and T.D. Anthopoulos, Adv. Funct. Mater. **27**, 1701818 (2017).

[43] I.T. Papadas, A. Savva, A. Ioakeimidis, P. Eleftheriou, G.S. Armatas, and S.A. Choulis, Mater. Today Energy **8**, 57 (2018).

[44] H. Zhang, H. Wang, H. Zhu, C. Chueh, W. Chen, S. Yang, and A.K.-Y. Jen, Adv. Energy Mater. **8**, 1702762 (2018).

[45] I.T. Papadas, A. Ioakeimidis, G.S. Armatas, and S.A. Choulis, Adv. Sci. **5**, 1701029 (2018).

[46] I. Zarazua, J. Bisquert, and G. Garcia-Belmonte, J. Phys. Chem. Lett. **7**, 525 (2016).

[47] J.-P. Correa-Baena, W. Tress, K. Domanski, E.H. Anaraki, S.-H. Turren-Cruz, B. Roose, P.P. Boix, M. Grätzel, M. Saliba, A. Abate, and A. Hagfeldt, Energy Environ.

Sci. **10**, 1207 (2017).

[48] D. Liu, S. Li, P. Zhang, Y. Wang, R. Zhang, H. Sarvari, F. Wang, J. Wu, Z. Wang, and Z.D. Chen, Nano Energy **31**, 462 (2017).

[49] B.-X. Chen, H.-S. Rao, W.-G. Li, Y.-F. Xu, H.-Y. Chen, D.-B. Kuang, and C.-Y. Su, J. Mater. Chem. A **4**, 5647 (2016).

[50] S.S. Shin, E.J. Yeom, W.S. Yang, S. Hur, M.G. Kim, J. Im, J. Seo, J.H. Noh, and S. Il Seok, Science (80-. ). **356**, 167 (2017).

[51] J.H. Kim, P.-W. Liang, S.T. Williams, N. Cho, C.-C. Chueh, M.S. Glaz, D.S. Ginger, and A.K.-Y. Jen, Adv. Mater. **27**, 695 (2015).

[52] W. Chen, Y. Wu, Y. Yue, J. Liu, W. Zhang, X. Yang, H. Chen, E. Bi, I. Ashraful, M. Gratzel, and L. Han, Science (80-. ). **350**, 944 (2015).

[53] M.-H. Liu, Z.-J. Zhou, P.-P. Zhang, Q.-W. Tian, W.-H. Zhou, D.-X. Kou, and S.-X. Wu, Opt. Express **24**, A1349 (2016).

[54] A.. Tavares, M.. da Silva Pereira, M.. Mendonça, M.. Nunes, F.. Costa, and C.. Sá, J. Electroanal. Chem. **449**, 91 (1998).

[55] A. Guerrero, G. Garcia-Belmonte, I. Mora-Sero, J. Bisquert, Y.S. Kang, T.J. Jacobsson, J.-P. Correa-Baena, and A. Hagfeldt, J. Phys. Chem. C **120**, 8023 (2016).

[56] H.-S. Kim, I.-H. Jang, N. Ahn, M. Choi, A. Guerrero, J. Bisquert, and N.-G. Park, J. Phys. Chem. Lett. **6**, 4633 (2015).

[57] J.-H. Kim, H.Y. Lee, and J.-Y. Lee, J. Nanosci. Nanotechnol. **18**, 2021 (2018).

[58] P.P. Boix, G. Garcia-Belmonte, U. Muñecas, M. Neophytou, C. Waldauf, and R. Pacios, Appl. Phys. Lett. **95**, 233302 (2009).

[59] A. Guerrero, J. You, C. Aranda, Y.S. Kang, G. Garcia-Belmonte, H. Zhou, J. Bisquert, and Y. Yang, ACS Nano **10**, 218 (2016).

[60] D. Zhao, M. Sexton, H.-Y. Park, G. Baure, J.C. Nino, and F. So, Adv. Energy Mater. **5**, 1401855 (2015).

[61] W. Shockley and W.T. Read, Phys. Rev. **87**, 835 (1952).

[62] R.N. Hall, Phys. Rev. **87**, 387 (1952).

[63] M.M. Mandoc, F.B. Kooistra, J.C. Hummelen, B. de Boer, and P.W.M. Blom, Appl. Phys. Lett. **91**, 263505 (2007).

[64] S.R. Cowan, A. Roy, and A.J. Heeger, Phys. Rev. B **82**, 245207 (2010).

[65] J.F. Marco, J.R. Gancedo, M. Gracia, J.L. Gautier, E. Ríos, and F.J. Berry, J. Solid State Chem. **153**, 74 (2000).

[66] T.J. Chuang, C.R. Brundle, and D.W. Rice, Surf. Sci. **59**, 413 (1976).

[67] A.C. Tavares, M.A.M. Cartaxo, M.I. da Silva Pereira, and F.M. Costa, J. Solid State Electrochem. **5**, 57 (2001).

[68] a. C. Tavares, M. a M. Cartaxo, M.I. Da Silva Pereira, and F.M. Costa, J. Electroanal. Chem. **464**, 187 (1999).

**FIGURES**

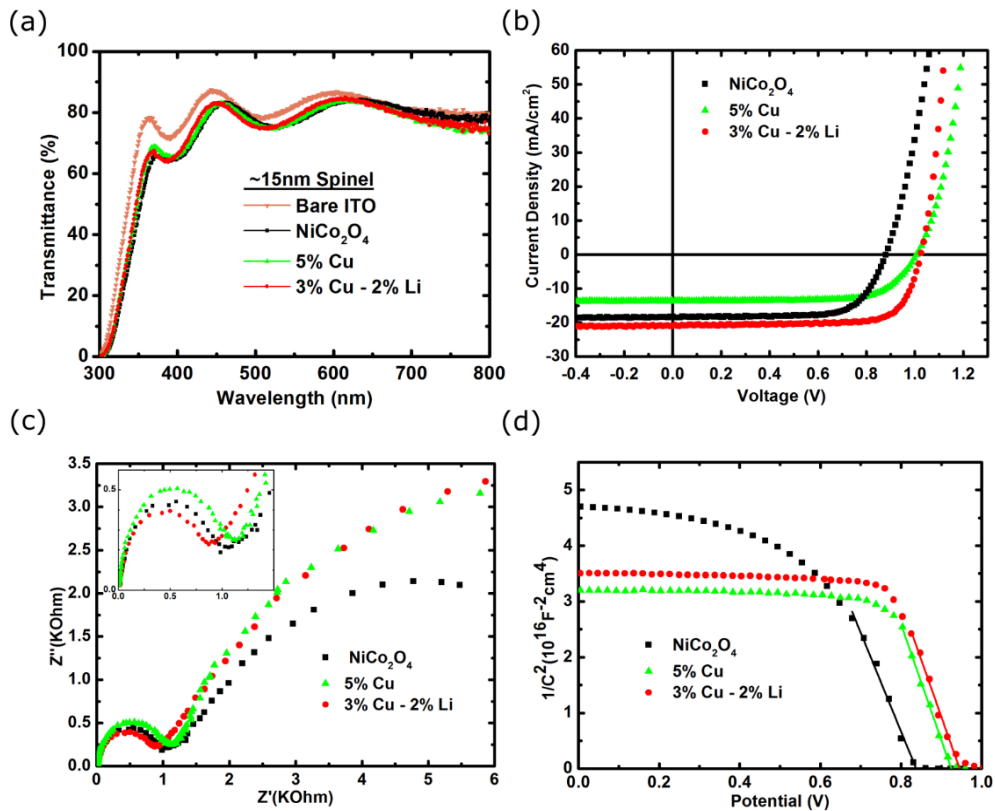

Fig. 1 (a) Transmittance measurements of bare glass/ITO and different type of doped 15 nm $NiCo_2O_4$ fabricated on glass/ITO substrates. (b) J-V curves, (c) Nyquist (inset: zoom-in at the high frequency region) and (d) Mott-Schottky plots of PVSC using 15 nm undoped, 5% Cu and 3% Cu – 2% Li doped $NiCo_2O_4$ HTL.

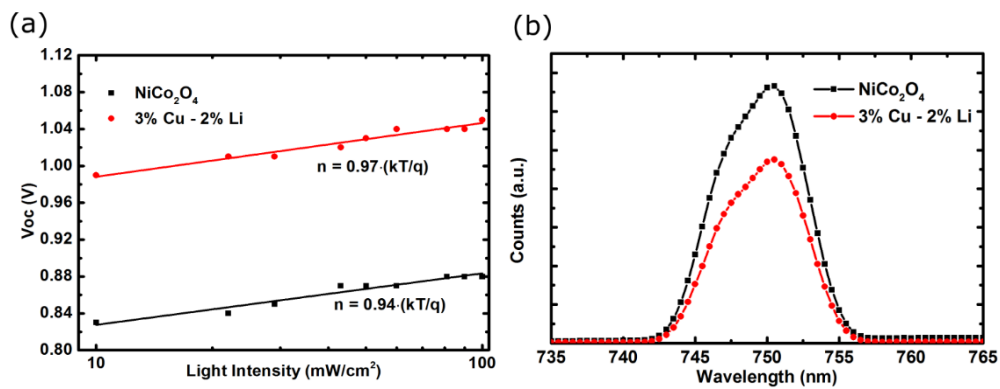

Fig. 2 (a) $V_{oc}$ – light intensity measurements of PVSC using 15 nm-sized undoped and 3% Cu – 2 % Li HTL. (b) Steady-state room temperature photoluminescence (PL) spectra of 350 nm thick perovskite films fabricated on 15 nm unmodified and 3% Cu – 2 % Li co-doped NiCo$_2$O$_4$ on glass/ITO substrate.

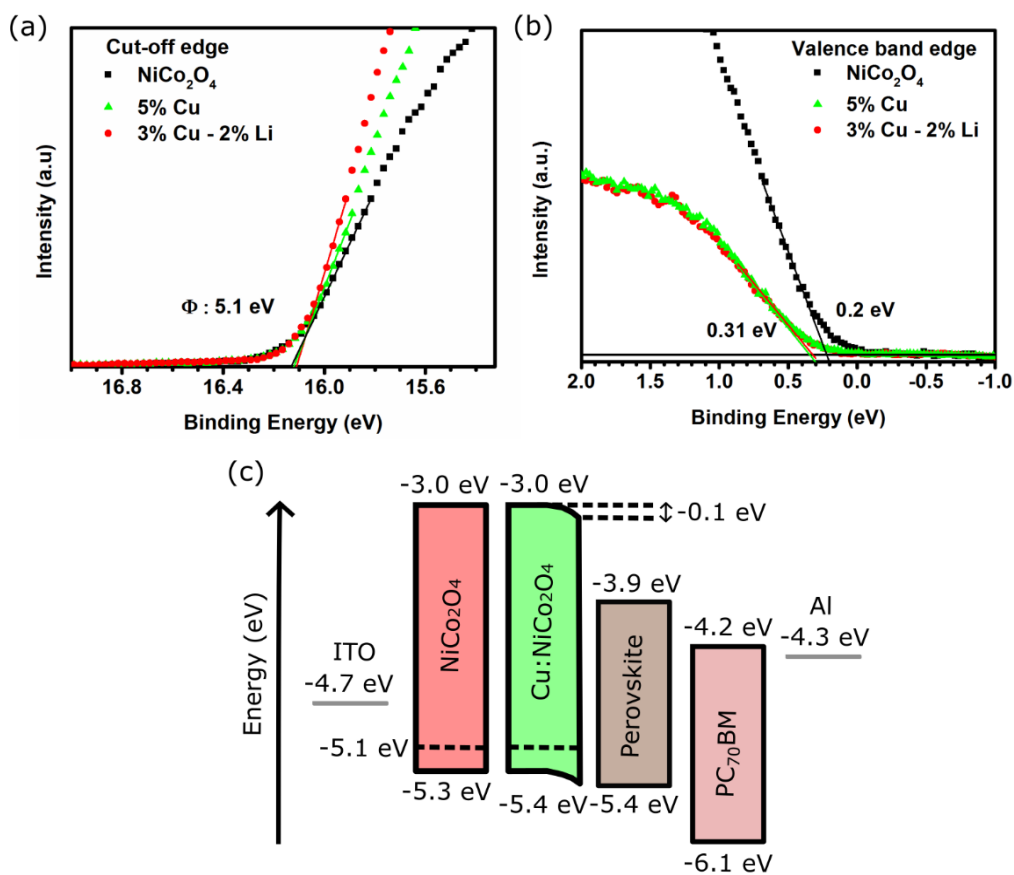

Fig. 3. (a) The high binding energy region and (b) valance band region near the Fermi level of the UPS spectra for undoped, 5% Cu and 3% Cu – 2% Li co-doped NiCo$_2$O$_4$ HTLs. (c) Schematic representation of energy band levels of the corresponding perovskite solar cells incorporating 5% Cu and 3% Cu – 2% Li doped NiCo$_2$O$_4$ (green bar) and undoped NiCo$_2$O$_4$ HTLs (red bar). In the case of the doped NiCo$_2$O$_4$ the band bending indicates the cathodic polarization effect at the surface region of the doped NiCo$_2$O$_4$ HTL.

# Supplementary Information

# Enhanced Photovoltaic Performance of Perovskite Solar Cells by Co-Doped Spinel Nickel Cobaltite Hole Transporting Layer

*Apostolos Ioakeimidis[1], Ioannis T. Papadas[1], Dimitris Tsikritzis[1], Gerasimos S. Armatas[2], Stella Kennou[3], Stelios A. Choulis[1]\**

[1] Molecular Electronics and Photonics Research Unit, Department of Mechanical Engineering and Materials Science and Engineering, Cyprus University of Technology, Limassol, 3603, Cyprus
[2] Department of Materials Science and Technology, University of Crete, Heraklion 71003, Greece
[3] Department of Chemical Engineering, University of Patras, 26504, Patras, Greece

*Corresponding Author: Prof. Stelios A. Choulis
E-mail:stelios.choulis@cut.ac.cy

# Materials and Methods

*Materials*:

Prepatterned glass-ITO substrates (sheet resistance 4Ω sq$^{-1}$) were purchased from Psiotec Ltd., Pb(CH$_3$CO$_2$)$_2$·3H$_2$O from Alfa Aesar, methylammonium iodide (MAI) and methylamonium bromide (MABr) from Dyenamo Ltd., PC[70]BM from Solenne BV. All the other chemicals used in this study were purchased from Sigma-Aldrich.

*Combustion Synthesis of NiCo$_2$O$_4$*:

For the combustion synthesis of NiCo$_2$O$_4$ NPs, 0.5 mmol Ni(NO$_3$)$_2$·6H$_2$O, 1 mmol Co(NO$_3$)$_2$·6H$_2$O, and tartaric acid were mixed in the 15 mL 2-methoxy ethanol solution For the preparation of the doped NiCo$_2$O$_4$ appropriate amounts of Cu(NO$_3$)$_2$·3H$_2$O and Li(CH$_3$CO$_2$)·2H$_2$O where added to the previous solution for 5 mol% Cu and 3 mol%

Cu – 2 mol% Li respectively. Then, 150 uL $HNO_3$ (69 wt% $HNO_3$) were added slowly into the mixture, and the solution stirred up to almost complete homogeneity. The whole solution was allowed under stirring for 30 min at 60 °C. The ratio of the total metal nitrates and tartaric acid was 1. Thereafter, the violet colored solution was used for the combustion synthesis of the $NiCo_2O_4$ NPs on the various substrates. Doctor blade technique was applied for the fabrication of the precursor films on the various substrates. The resulting light violet colored films were dried at 100 °C for 30 min and used as a precursor for the combustion synthesis of $NiCo_2O_4$ NPs. Subsequently, the obtained films were heated at 250 °C in ambient atmosphere for 1 h in a preheated oven to complete the combustion process and then left to cool down at room temperature.

*Device Fabrication*:

The inverted solar cells under study was ITO/$NiCo_2O_4$-NPs/$CH_3NH_3PbI_3$/PC[70]BM/Al. ITO substrates were sonicated in acetone and subsequently in isopropanol for 10 min and heated at 100 °C on a hot plate 10 min before use. The perovskite solution was prepared 30 min prior spin coating by mixing $Pb(CH_3CO_2)_2 \cdot 3H_2O$:methylamonium iodide (1:3) at 40 wt% in dimethylformamide (DMF) with the addition of 1.5% mole of MABr. The precursor was filtered with 0.1 µm polytetrafluoroethylene (PTFE) filters. The perovskite precursor solution was deposited on the HTLs by static spin coating at 4000 rpm for 60 s and annealed for 5 min at 85 °C, resulting in a film with a thickness of ~350 nm. The PC[70]BM solution, 20 mg mL$^{-1}$ in chlorobenzene, was dynamically spin coated on the perovskite layer at 1000 rpm for 30 s. Finally, 100 nm Al layers were thermally evaporated through a shadow mask to finalize the devices giving an active area of 0.9 mm$^2$. Encapsulation was applied directly after evaporation in the glove box using a glass coverslip and an Ossila E131 encapsulation epoxy resin activated by 365 nm UV irradiation.

*Characterization*:

For UV–vis absorption, XRD, four-point probe and AFM measurements the films were fabricated on quartz substrates. Transmittance, AFM, XPS and UPS measurements were conducted on films fabricated on glass/ITO substrates. XRD patterns were collected on a PANanalytical X'pert Pro MPD powder diffractometer (40 kV, 45 mA) using Cu Kα radiation (λ = 1.5418 Å). Transmittance and absorption measurements were performed with a Schimadzu UV-2700 UV–vis spectrophotometer. The thickness of the films was measured with a Veeco Dektak 150 profilometer. The current density–voltage (*J/V*) and Voc-intensity were obtained using a Botest LIV Functionality Test System measured with 10 mV voltage steps and 40 ms of delay time. For illumination, a calibrated Newport Solar simulator equipped with a Xe lamp was used, providing an AM1.5G spectrum at 100 mW cm$^{-2}$ as measured by a certified oriel 91150 V calibration cell. A shadow mask was attached to each device prior to measurements to accurately define the corresponding device area. Steady-state PL experiments were performed on a Fluorolog-3 Horiba Jobin Yvon spectrometer based on an iHR320 monochromator equipped with a visible photomultiplier tube (Horiba TBX-04 module). The PL was non-resonantly excited at 550 nm with the line of a 5 mW Oxxius laser diode. AFM images were obtained using a Nanosurf easy scan 2 controller under the tapping mode. Electrical conductivity measurements were performed using a four-point microposition probe, Jandel MODEL RM3000. EIS and MS measurements were performed using a Metrohm Autolab PGSTAT 302N, where for the EIS a red light-emitting diode (at 625 nm) was used as the light source calibrated to 100 mW cm$^{-2}$. For EIS a small AC perturbation of 20 mV was applied to the devices, and the different current output was measured throughout a frequency range of 1 MHz to 1 Hz. The steady state DC bias was kept at 0 V throughout the EIS experiments. X-ray photoelectron spectra (XPS) and Ultraviolet Photoelectron Spectra (UPS) were recorded by using a Leybold EA-11 electron analyzer operating in constant energy mode at a pass energy of 100 eV and at a constant retard ratio of 4 eV for XPS and UPS, respectively. The spectrometer energy scale was calibrated by the Au $4f_{7/2}$ core level binding energy, BE, (84.0 ± 0.1 eV) and the energy scale of the UPS measurements was referenced to the Fermi level position of Au at a binding energy of 0 eV. All binding energies were referred to the C 1s peak at 284.8 eV of surface adventitious carbon. The X-ray source for all measurements was a non-monochromatized Al Kα line at 1486.6 eV (12 keV with 20 mA anode current). For UPS measurements, the He I (21.22 eV) excitation line was used. A negative bias of 12.22 V was applied to the samples during UPS measurements in order to separate secondary electrons originating from the sample and the spectrometer. The sample work function was determined by subtracting the high binding energy cut-off from the He I excitation energy (21.22 eV). The position of the high-energy cut-off was determined by the intersection of a linear fit of the high binding portion of the spectrum with the background. Similarly, the valence band maximum is determined with respect to the Fermi level, from the linear extrapolation of the valence band edge to the background.

# Supplementary Information Figures

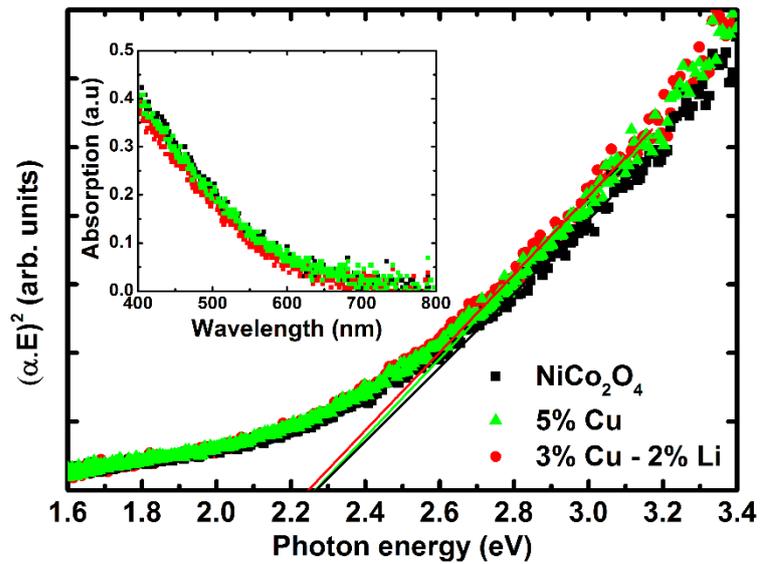

Figure S1. Tauc-plot of undoped (black rectangles), 5% Cu (green triangles) and 3% Cu – 2% Li (red circles) NiCo$_2$O$_4$ films. The inset shows the absorption measurements of the corresponding films.

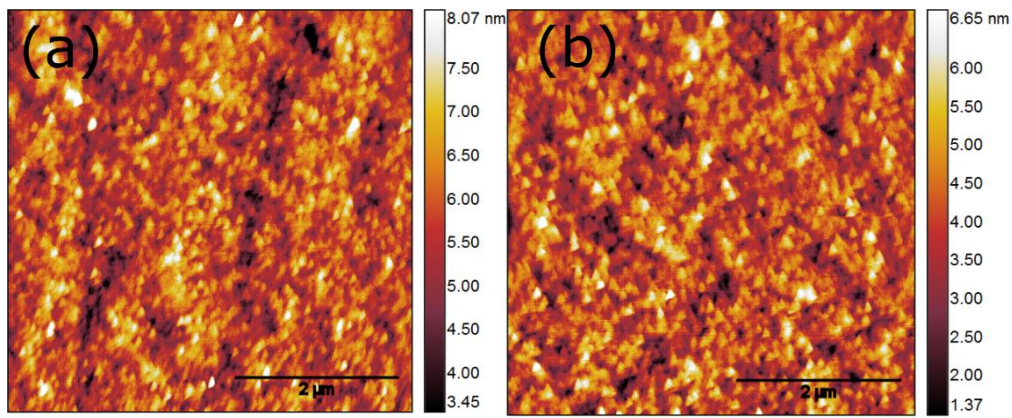

Figure S2. AFM topography images of (a) 5% Cu (b) 2% Li- 3% Cu doped NiCo2O4 films, respectively, fabricated on quartz substrate. The films exhibit a roughness of (a) 0.7, (b) 0.8 nm.

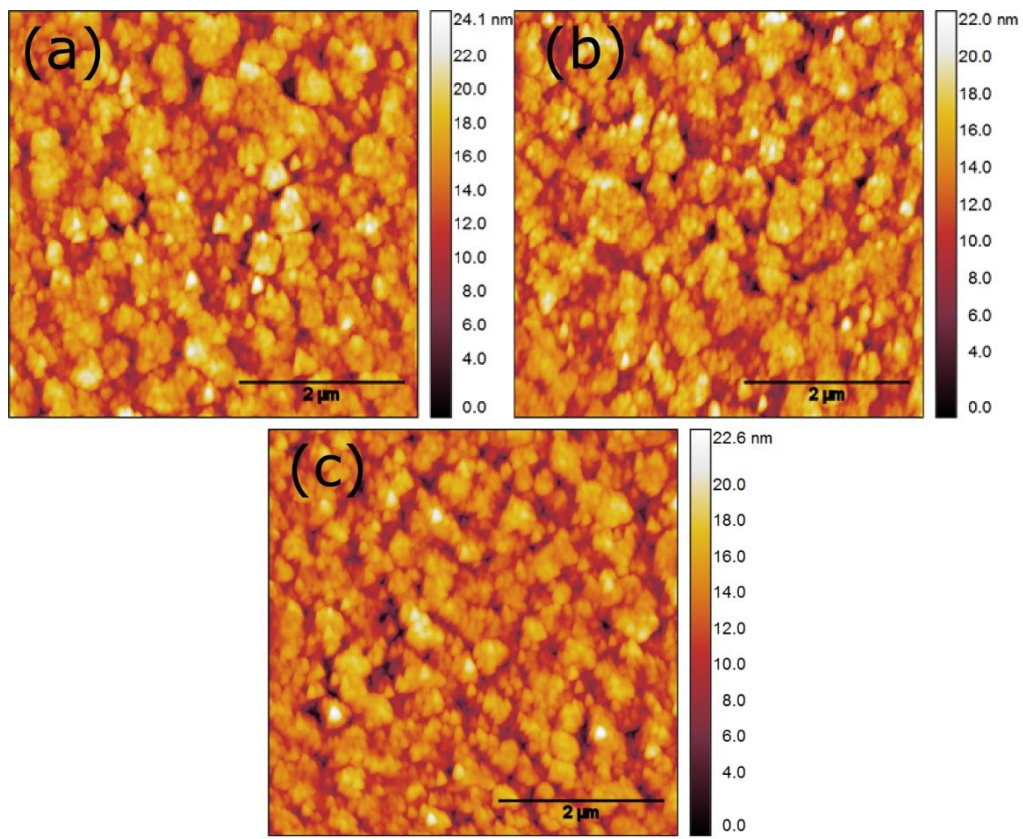

Figure S3. AFM topography images of (a) 5% Cu and (b) 3% Cu - 2% Li doped NiCo$_2$O$_4$ films, respectively, fabricated on (c) glass/ITO substrate. The films exhibit similar roughness (a) 2.9, (b) 3.0 and (c) 2.8 nm.

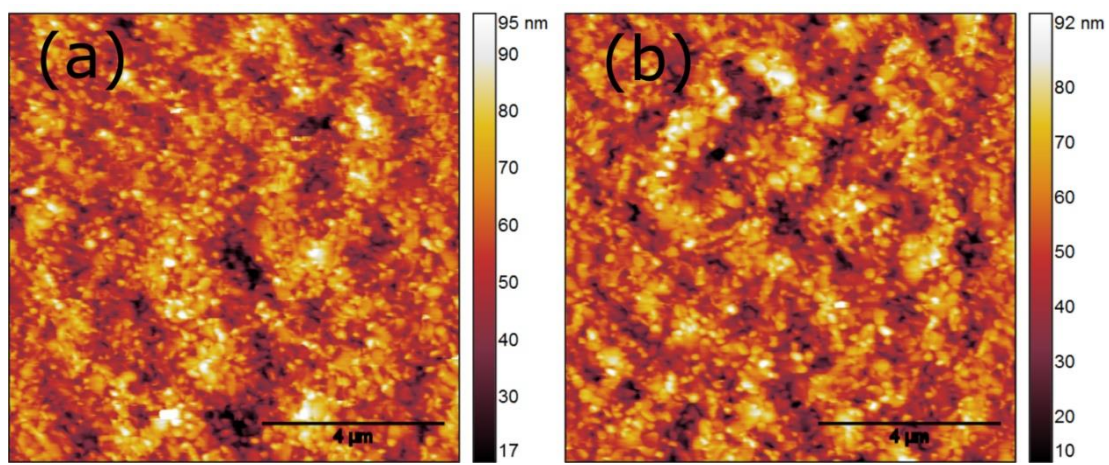

Figure S4. AFM topography images (10x10 μm) of perovskite active layers on (a) undoped and (b) 3% Cu – 2 % Li doped NiCo₂O₄ 350 nm thick perovskite.

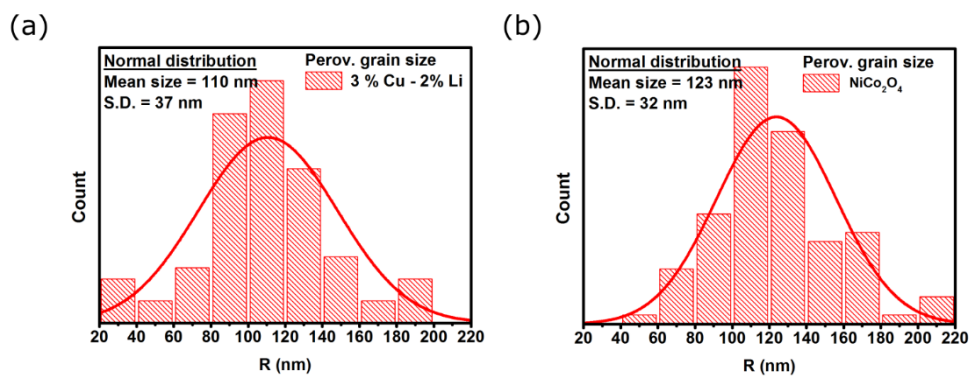

Figure S5. Distribution of perovskite grain size and the extracted parameters of mean value and standard deviation using normal distribution fit curves for perovskite films on undoped and 3% Cu – 2 % Li doped NiCo₂O₄, for 350 nm thick perovskite.

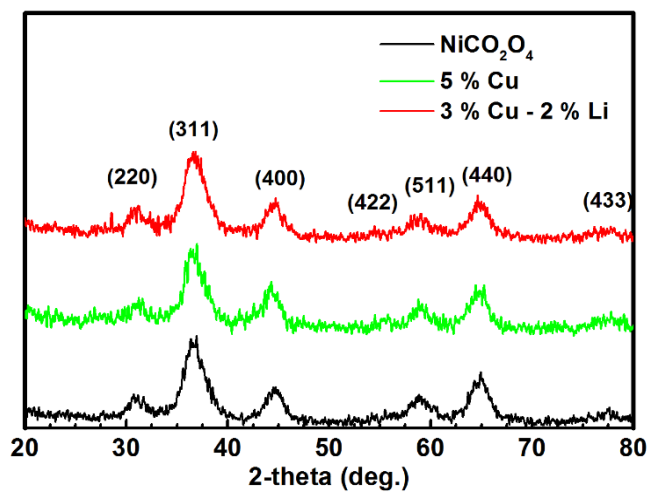

Figure S6. X-ray diffraction (XRD) patterns of undoped, 5% Cu doped NiCo₂O₄ 5, 3% Cu – 2% Li co-doped NiCo₂O₄ powder.

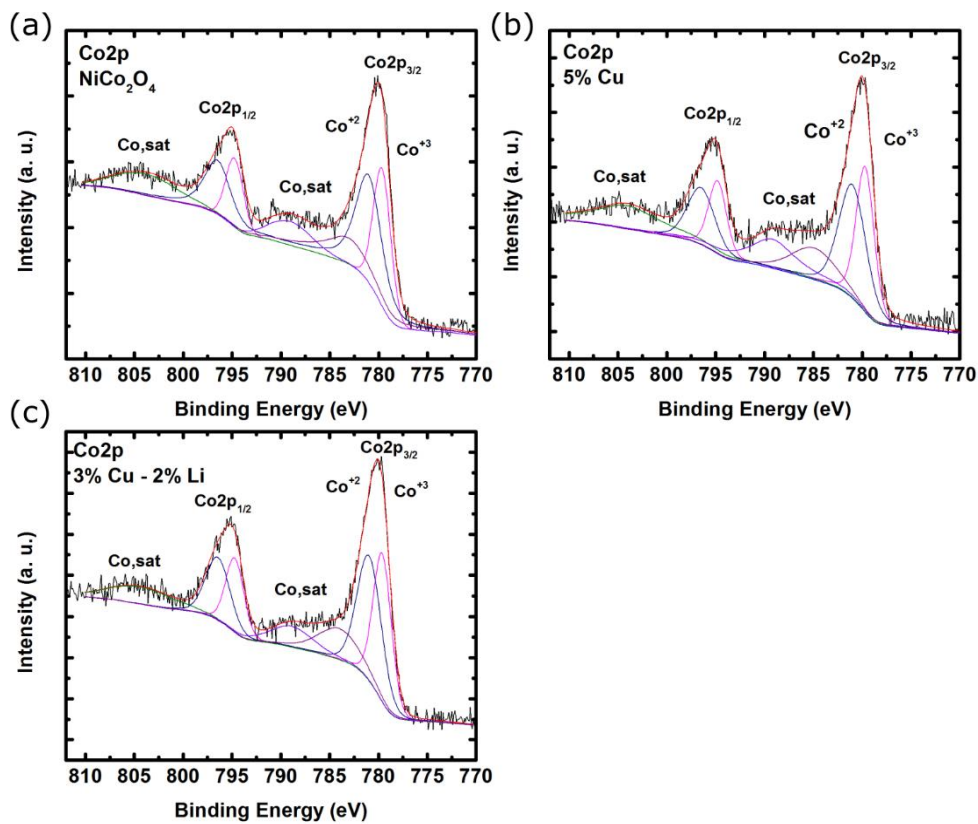

Figure S7. XPS spectra of the Co2p for (a) undoped, (b) 5% Cu doped $NiCo_2O_4$, (c) 3% Cu – 2 % Li co-doped $NiCo_2O_4$ films.

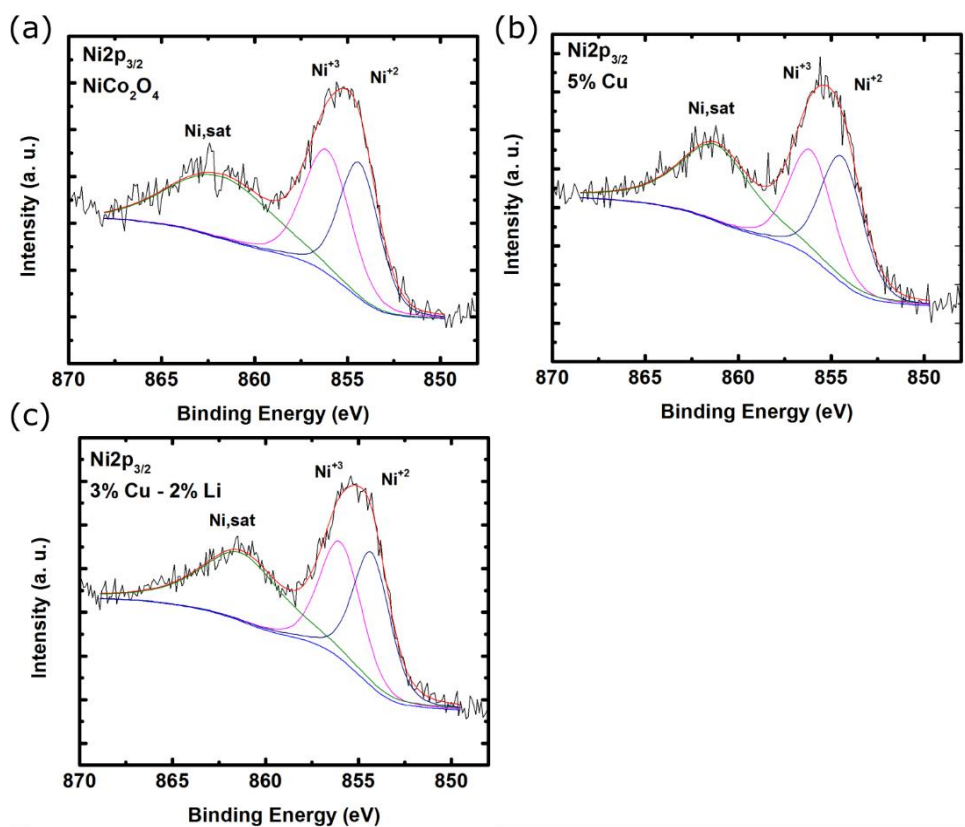

Figure S8. XPS spectra of the Ni2p$_{3/2}$ for (a) undoped, (b) 5% Cu doped NiCo$_2$O$_4$ (c) 3% Cu – 2 % Li co-doped NiCo$_2$O$_4$ films

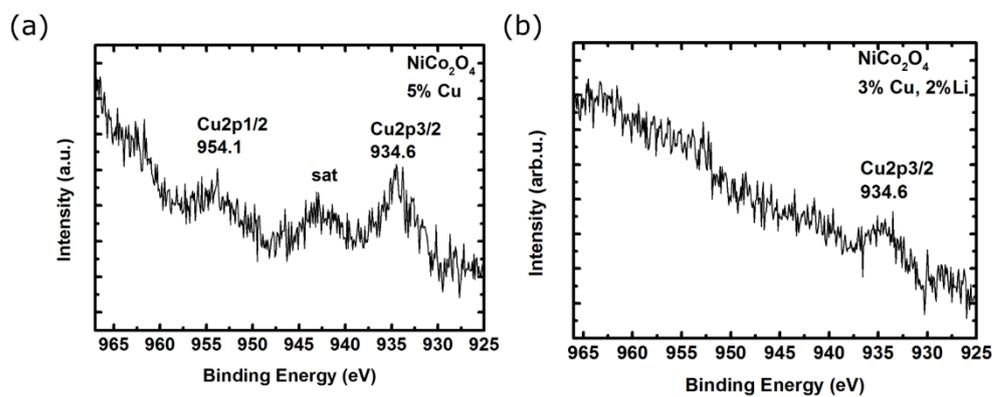

Figure S9. XPS spectra of the Cu2p for (a) 5% Cu doped NiCo$_2$O$_4$ (b) 3% Cu – 2 % Li co-doped NiCo$_2$O$_4$ films.